\documentclass{ws-p8-50x6-00}

\input{psfig}

\begin{document}

\title{\vspace*{-1.0cm} \hfill {\rm MKPH-T-00-22}\\ \vspace{0.5cm}
%\title{
Helicity Amplitudes and Sum Rules for Real and Virtual Photons}

\author{L. TIATOR}

\address{Institut f\"{u}r Kernphysik, Johannes Gutenberg-Universit\"{a}t,
J.~J.~Becher-Weg 45,\\
D-55099 Mainz, Germany\\
E-mail: tiator@kph.uni-mainz.de}

\maketitle

\abstracts{Results of the recently developed unitary isobar model
(MAID) are presented for helicity amplitudes, spin asymmetries,
structure functions and relevant sum rules for real and virtual
photons in the resonance region. Our evaluation of the
energy-weighted integrals is in good agreement for the proton but
shows big discrepancies for the neutron. }

\section{Introduction}\label{sec:intro}
The spin structure of the nucleon in the resonance region is of
particular interest to understand the rapid transition from resonance
dominated coherent processes to incoherent processes of deep
inelastic scattering (DIS) off the constituents. Here we present the
results of the recently developed unitary isobar model
(MAID)\cite{Dre98a} for the spin asymmetries, structure functions and
relevant sum rules in the resonance region. This model describes the
presently available data for single-pion photo- and electroproduction
up to a total $cm$ energy $W_{\rm max}= 2$ GeV and for $Q^2\le$ 4
(GeV/c)$^2$. It is based on effective Lagrangians for Born terms and
vector meson exchange (background) and resonance contributions
modeled by Breit-Wigner functions. All major resonances below
$W=1700$ MeV are included. The respective multipoles are constructed
in a gauge-invariant and unitary way for each partial wave.  The eta
production is included in a similar way\cite{Kno95}, while the
contribution of more-pion and higher channels is modeled by
comparison with the total cross sections and simple phenomenological
assumptions.

\section{Formalism}\label{sec:form}

The differential cross section for exclusive electroproduction of
mesons from polarized targets using polarized electrons, e.g. $\vec
p(\vec e,e'\pi^0)p$ can be parametrized in terms of 18 response
functions\cite{Dre92}, a total of 36 is possible if in addition also
the recoil polarization is observed. Due to the azimuthal symmetry
most of them vanish by integration over the angle $\phi$ and only 5
total cross sections remain. The differential cross section for the
electron is then given by
\begin{equation}
\frac{d\sigma}{d\Omega\ dE'} = \Gamma\sigma (\nu,Q^2)\,,
\label{eq1}
\end{equation}
\begin{equation}
\sigma =\sigma_T+\epsilon\sigma_L+
P_y\sqrt{2\epsilon(1+\epsilon)}\
         \sigma_{LT}+
         hP_x\sqrt{2\epsilon(1-\epsilon)}\
         \sigma_{LT'}+hP_z\sqrt{1-\epsilon^2}\sigma_{TT'}\, ,
\label{eq2}
\end{equation}
where $\Gamma$ is the flux of the
virtual photon field and the $\sigma_i,$ $i=L$, $T$, $LT$, $LT'$,
$TT'$, are functions of the $lab$ energy of the virtual photon
$\nu$ and the squared four-momentum transferred $Q^2$. These
response functions can be separated by varying the transverse
polarization $\epsilon$ of the virtual photon as well as the
polarizations of the electron ($h$) and proton ($P_z$ parallel,
$P_x$ perpendicular to the virtual photon, in the scattering plane
and $P_y$ perpendicular to the scattering plane). In particular,
$\sigma_{T}$ and $\sigma_{TT'}$ can be expressed in terms of the
total cross sections for excitation of hadronic states with spin
projections $3/2$ and $1/2$:
$\sigma_{T}=(\sigma_{3/2}+\sigma_{1/2})/2$ and
$\sigma_{TT'}=(\sigma_{3/2}-\sigma_{1/2})/2.$

Here we use Hand's notation with the equivalent photon $cm$ energy
$K=(W^2-m^2)/(2W)$ for the virtual photon flux. Correspondingly,
the phase space factors of the cross sections are given by $q/K$,
where $q$ is the pion momentum in the $cm$.

In inclusive electron scattering  $\vec e+\vec N\rightarrow X$,
only 4 cross sections $\sigma_T$, $\sigma_L$, $\sigma_{LT'}$ and
$\sigma_{TT'}$ appear, the fifth cross section, $\sigma_{LT}$,
vanishes due to unitarity when all open channels are summed up.
The individual channels, however, give finite contributions.

The Gerasimov-Drell-Hearn (GDH) sum rule is only derived for real
photons. It is based on unitarity and low-energy theorems and the
assumption of the convergence of an unsubtracted dispersion
relation,
\begin{eqnarray}
I_{GDH} &=& \frac{m^2}{8\pi^2\alpha}\int_{\nu_0}^{\infty}
            \left (\sigma_{1/2}-\sigma_{3/2}
            \right )\ \frac{d\nu}{\nu}\
            = -\frac{\kappa^2}{4} \,.
\label{eq3}
\end{eqnarray}
This sum rule is often presented without the leading factor in
front of the integral, the numerical conversion is
$8\pi^2\alpha/m^2=254.8\mu b$.

It can be generalized in various ways. Three forms often used in
the literature are summing up only contributions from
$\sigma_{TT'}$ with no longitudinal terms,
\begin{eqnarray}
I_{GDH}^{(a)}(Q^2)
&=&\frac{m^2}{8\pi^2\alpha}\int_{\nu_0}^{\infty}
           (1-x)
           \left (\sigma_{1/2}-\sigma_{3/2}
           \right )\ \frac{d\nu}{\nu}\ \,,
\label{eq4}
\\
I_{GDH}^{(b)}(Q^2)
&=&\frac{m^2}{8\pi^2\alpha}\int_{\nu_0}^{\infty}
           \frac{K}{\mid\vec{k}_{\gamma}\mid}
           \left (\sigma_{1/2}-\sigma_{3/2}
           \right )\ \frac{d\nu}{\nu}\ \,,
\label{eq5}
\\
I_{GDH}^{(c)}(Q^2)
&=&\frac{m^2}{8\pi^2\alpha}\int_{\nu_0}^{\infty}
           \left (\sigma_{1/2}-\sigma_{3/2}
           \right )\ \frac{d\nu}{\nu}\ \,.
\label{eq6}
\end{eqnarray}
The factor $K/\mid\vec{k}_{\gamma}\mid$ can also be expressed as
$(1-x)/\sqrt{1+\gamma^2}$. The relations between the $\sigma_i$ and
the quark structure functions $g_1$ and $g_2$ can be read off the
following equations, which define further possible generalizations of
the GDH integral\cite{Ger65} and the Burkhardt-Cottingham (BC) sum
rule\cite{Bur70}, which in addition also include
longitudinal-transverse interference terms,
\begin{eqnarray}
I_1(Q^2) &=& \frac{2m^2}{Q^2}\int_{0}^{x_0}g_1(x,Q^2)\ dx
\nonumber
\\
&=&\frac{m^2}{8\pi^2\alpha}\int_{\nu_0}^{\infty}
           \frac{1-x} {1+\gamma^2}
           \left (\sigma_{1/2}-\sigma_{3/2}
           -2\gamma\,\sigma_{LT'}\right )\ \frac{d\nu}{\nu}\ \,,
\label{eq7}
\\
I_2(Q^2) &=& \frac{2m^2}{Q^2}\int_{0}^{x_0}g_2(x,Q^2)\ dx \nonumber
\\
&=&  \frac{m^2}{8\pi^2\alpha}\int_{\nu_0}^{\infty}
           \frac{1-x} {1+\gamma^2}
           \left (\sigma_{3/2}-\sigma_{1/2}
           -\frac{2}{\gamma}\,\sigma_{LT'}\right )\ \frac{d\nu}{\nu} \,,
\label{eq8}
\\
I_3(Q^2) &=& \frac{2m^2}{Q^2}\int_{0}^{x_0}(g_1(x,Q^2)+g_2(x,Q^2))\ dx \nonumber
\\
&=&  -\frac{m^2}{4\pi^2\alpha}\int_{\nu_0}^{\infty}
           \frac{1-x}{Q}\,\sigma_{LT'}\ d\nu\ = I_1+I_2 \,,
\label{eq9}
\end{eqnarray}
where $\gamma=Q/\nu$ and $x=Q^2/2m\nu$ the Bjorken scaling
variable, with $x_0$ ($\nu_0$) referring to the inelastic
threshold of one-pion production. Since $\sigma_{LT'}={\cal
O}(Q)$, the real photon limit of the integral $I_1$  is given by
the GDH sum rule $I_1(0)=I_{GDH}(0)=-\kappa_N^2/4,$ with
$\kappa_N$ the anomalous magnetic moment of the nucleon. At large
$Q^2$ the structure functions should depend only on $x,$ i.e.
$I_1\rightarrow 2m\Gamma_1/Q^2$ with $\Gamma_1=\int
g_1(x){\rm d}x=$ const. In the case of the proton, all experiments
for $Q^2> 1$GeV$^2$ yield $\Gamma_1>0.$ Therefore, a strong
variation of $I_1(Q^2)$ with a zero-crossing at $Q^2<1$ GeV$^2$ is
required in order to reconcile the GDH sum rule with the
measurements in the DIS region. The $I_2$ integral of
Eq.~(\ref{eq8}) is constrained by the BC sum rule, which requires
that the inelastic contribution for $0 < x <x_0 $ equals the
negative value of the elastic contribution, i.e.
\begin{equation}
I_2(Q^2) = \frac{2m^2}{Q^2}\int_{0}^{x_0}g_2(x,Q^2)\ dx
=\frac{1}{4}\frac{G_M(Q^2)-G_E(Q^2)}{1+Q^2/4m^2}\,G_M(Q^2)\,,
\label{eq10}
\end{equation}
where $G_M$ and $G_E$ are the magnetic and electric Sachs form
factors respectively. At large $Q^2$ the integral vanishes as
$Q^{-10}$, while at the real photon limit
$I_2(0)=\kappa_N^2/4+e_N\kappa_N/4$, the two terms on the right
hand side corresponding to the contributions of $\sigma_{TT'}$ and
$\sigma_{LT'}$ respectively. Finally, Eq. (\ref{eq9}) defines an
integral $I_3(Q^2)$ as the sum of $I_1(Q^2)$ and $I_2(Q^2)$ and is
given by the unweighted integral over the longitudinal-transverse
interference cross section $\sigma_{LT'}$. At the real photon
point this integral is given by the GDH and BC sum rules,
$I_3(0)=e_N\kappa_N/4$. In particular this vanishes for the
neutron target.

\section{Unitary Isobar Model}\label{sec:uim}
Our calculation for the response functions $\sigma_i$ is based on the
Unitary Isobar Model (UIM) for one-pion photo- and electroproduction
of Ref.\cite{Dre98a}, accessible in the internet as the MAID program.
The model is constructed with effective Lagrangians for Born terms,
vector meson exchange in the $t$ channel (background), and the
dominant resonances up to the third resonance region are modeled
using Breit-Wigner functions with energy-dependent widths. For each
partial wave the multipoles satisfy gauge invariance and unitarity.
As in any realistic model a special effort is needed to describe the
$s$-channel multipoles $S_{11}$ and $S_{31}$. Even close at threshold
these multipoles pick up sizeable imaginary parts that cannot be
explained by nucleon resonances. In fact the $S_{11}(1535)$,
$S_{11}(1650)$ and the $S_{31}(1620)$ play only a minor role for the
complex phase of the $E_{0+}$ multipoles even at higher energies. The
main effect arises from pion rescattering. This we can take into
account by $K$-matrix unitarization. Furthermore we introduce a gauge
invariant contact term proportional to the anomalous magnetic moment
of the nucleon $\kappa_N$,
\begin{equation}
j_\kappa^\mu = \frac{ieg}{2m}\kappa_N
F(q_0^2)\frac{\sigma^{\mu\nu}k_\nu}{2m}\gamma_5 \,.
\label{eq11}
\end{equation}
The form factor $F(q_0^2)$, $q_0$ being the asymptotic pion
momentum, vanishes at threshold, consistent with chiral symmetry,
but gives rise to a cancellation of unphysically high momentum
components in the Born terms at high energies.

Due to unitarity each partial wave has to fulfill Watson's
theorem,
\begin{eqnarray}
t_{\gamma,\pi}^\alpha &=& t_{\gamma,\pi}^\alpha(background) +
t_{\gamma,\pi}^\alpha(resonances)\\ \nonumber &=& \pm \mid
t_{\gamma,\pi}^\alpha \mid e^{i\delta_{\pi N}^\alpha} \, .
\label{eq12}
\end{eqnarray}
In an isobar model this condition has to be constructed
explicitly. In Maid98 the background is real (except for the
S-waves) and a phase is added to the resonance. In Maid2000 both
background and resonance contributions are unitarized separately
for all partial waves up to $l=3$ in the following way
\begin{equation}
   t_{\gamma,\pi}^\alpha = t_{\gamma,\pi}^\alpha(Born+\omega,\rho)
   (1\,+\,i\,t_{\pi N}^I) +
   t_{\gamma,\pi}^\alpha(resonances)e^{i\Psi^\alpha}\, .
\label{eq13}
\end{equation}

\begin{figure}[ht]
\centerline{ \psfig{file=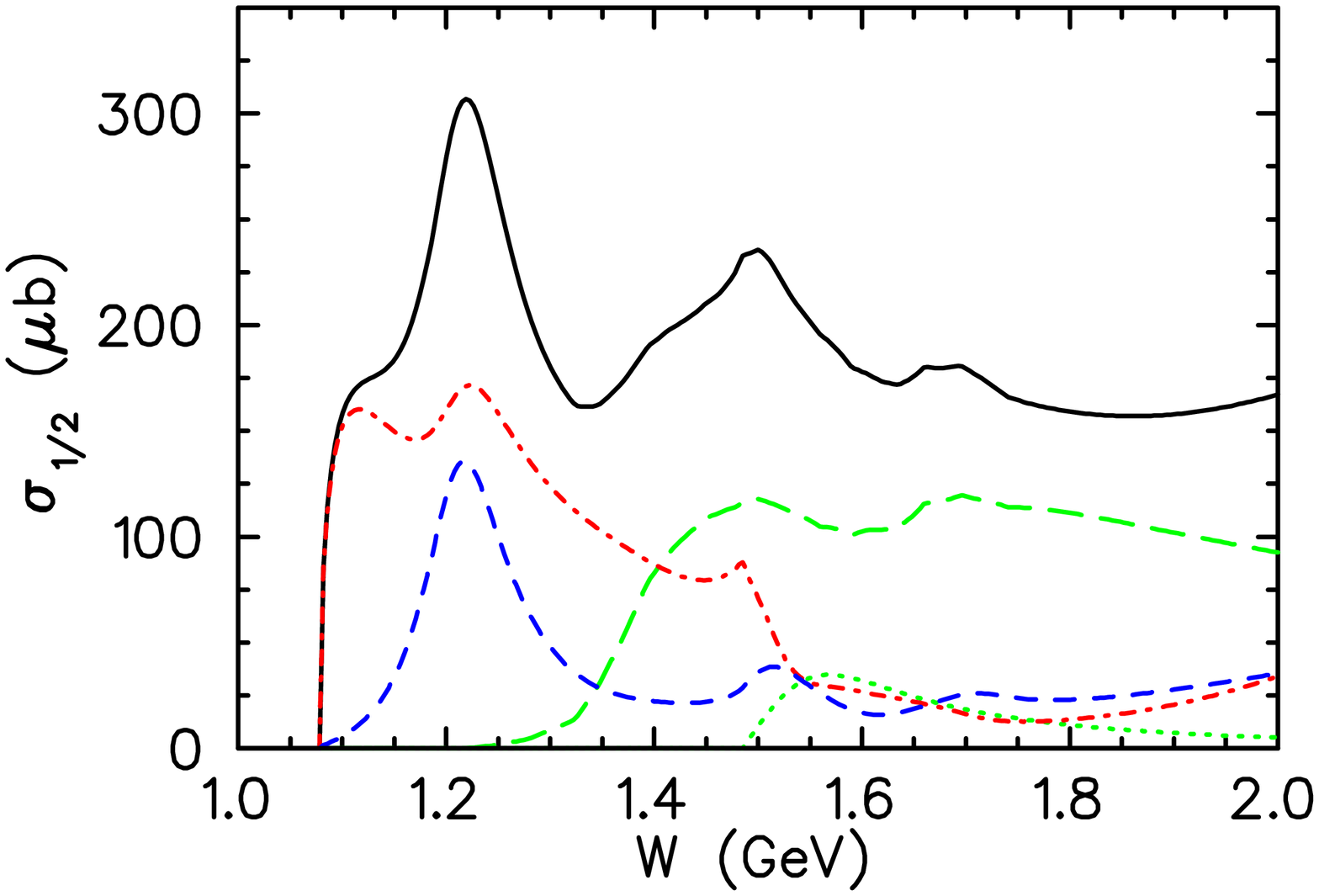,height=5.5cm,angle=90,silent=}
\hspace{0.1cm} \psfig{file=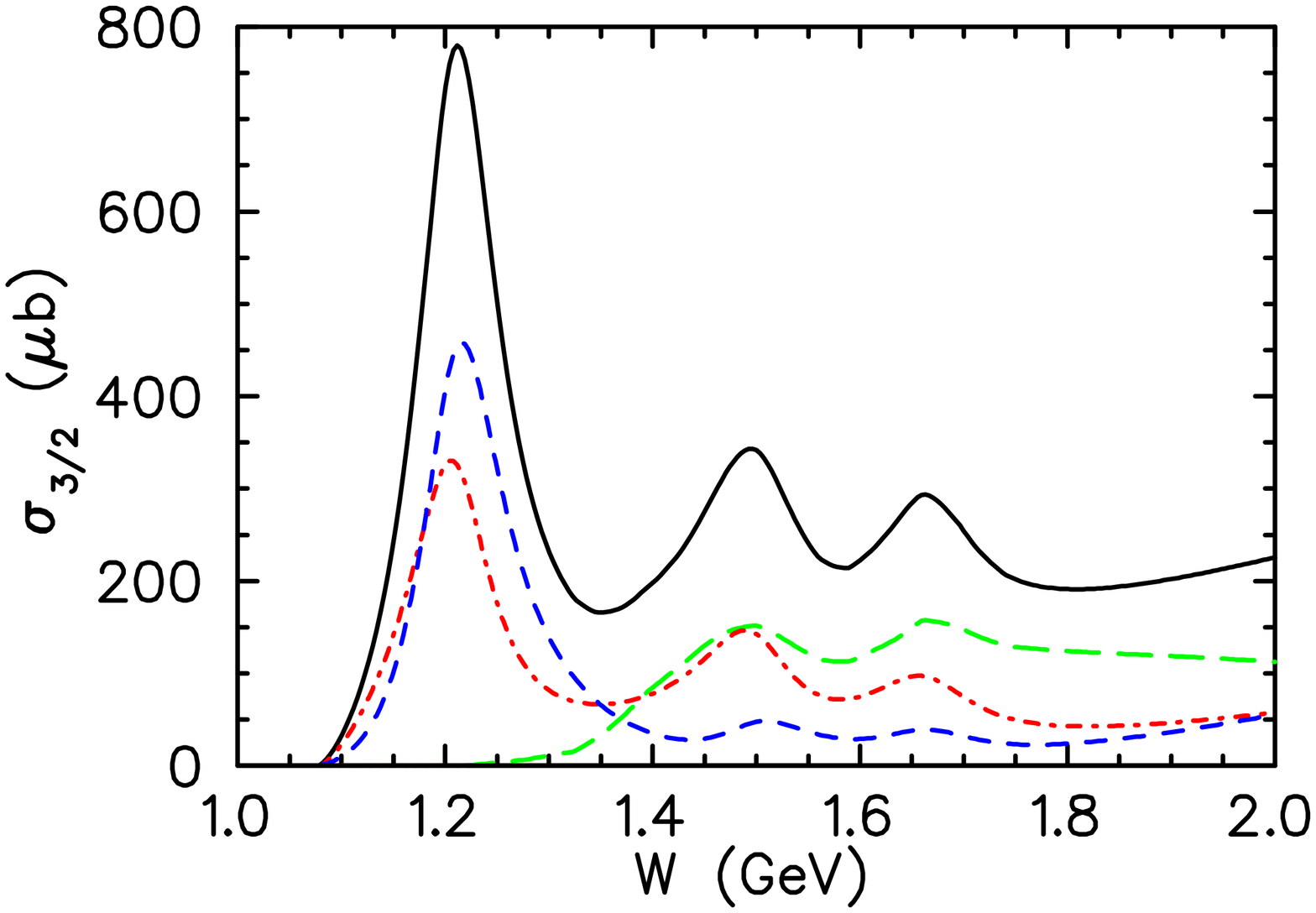,height=5.5cm,angle=90,silent=}
} \vspace{0.1cm} \centerline{
\psfig{file=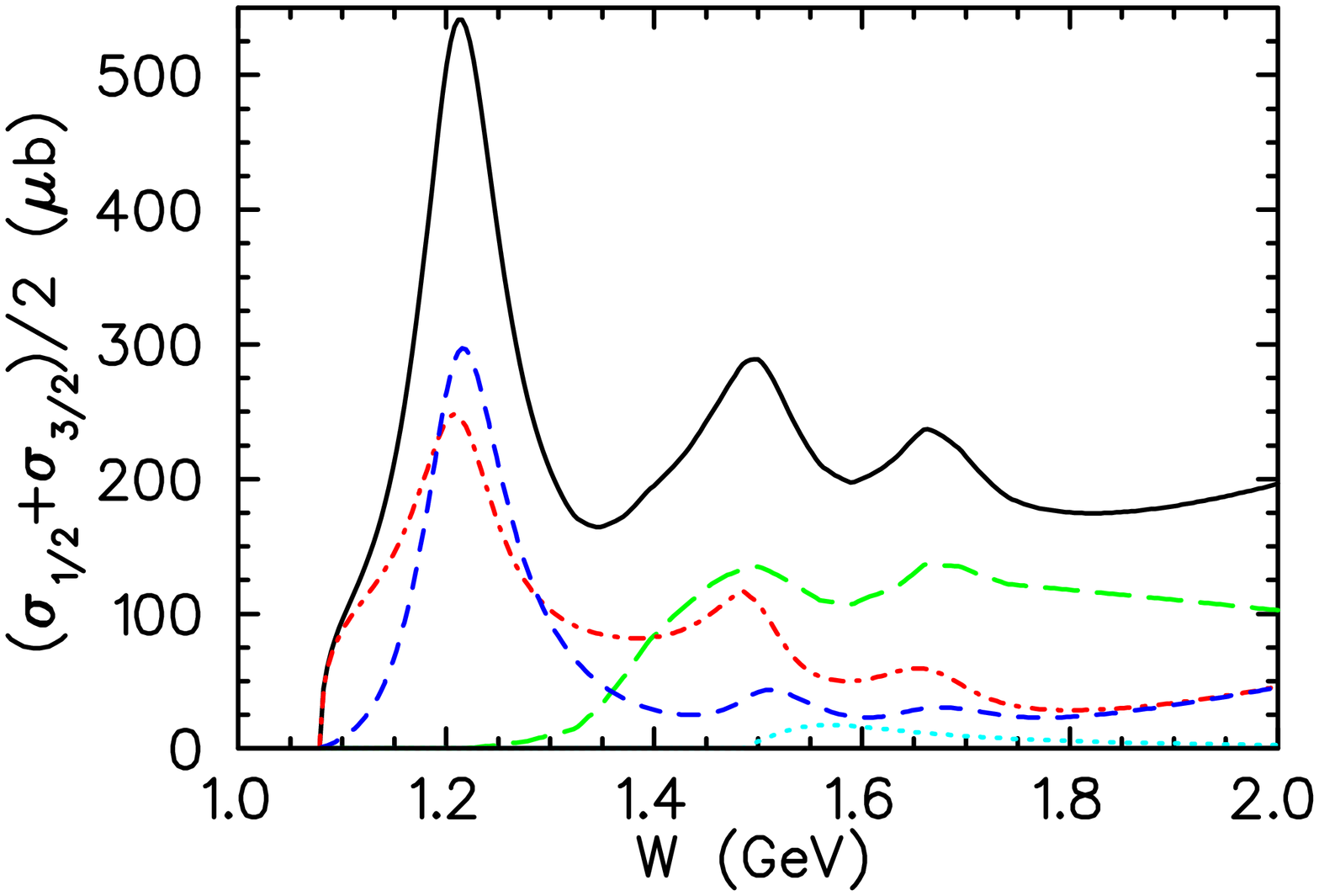,height=5.5cm,angle=90,silent=} \hspace{0.1cm}
\psfig{file=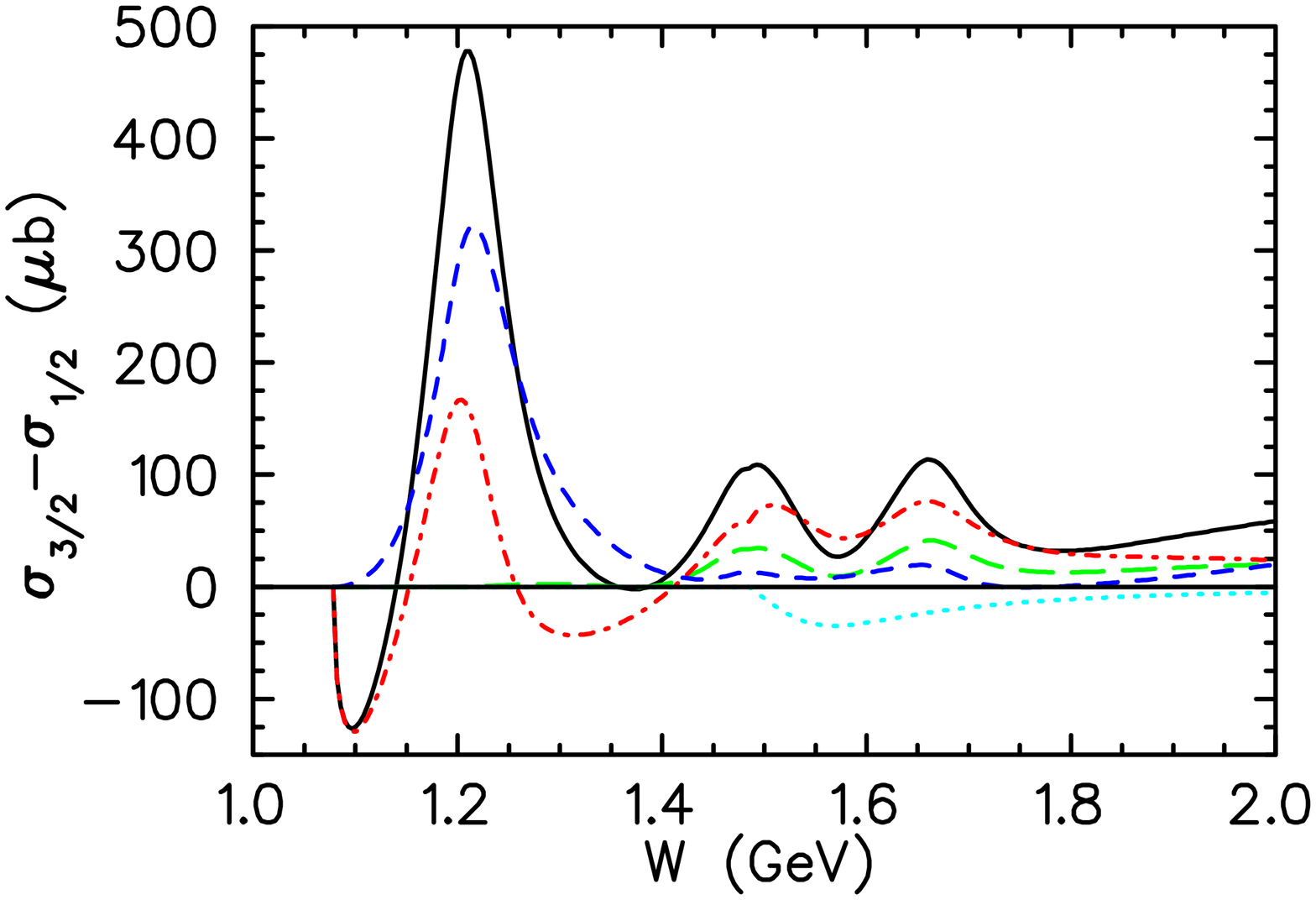,height=5.5cm,angle=90,silent=} } \vskip
-0.3cm \caption{\label{fig1}
  Helicity dependent cross sections for photoproduction on the proton target.
  The dashed, dash-dotted, long dashed and dotted lines show our
  calculations for $\pi^0$, $\pi^+$, two-pion and the $\eta$ cross
  sections, respectively. The solid curves give the total
  photoabsorption and contain the sum of all.}
\end{figure}

The UIM is able to describe the single-pion electroproduction channel
quite well. However, at higher energies the contributions from other
channels become increasingly important. In the structure functions
$\sigma_{T}$ and $\sigma_{TT'}$ we account for the $\eta$ and the
multi-pion production contributions extracting the necessary
information from the existing data for the total cross
section\cite{Dre98b}. In Fig.~\ref{fig1} we show the individual
channels for the total helicity dependent cross sections
$\sigma_{1/2}$, $\sigma_{3/2}$, $\sigma_T$ and
$\Delta\sigma=2\sigma_{TT'}$ at $Q^2=0$. Due to the non-regularized
Born terms in the $1\pi$ channels the cross sections start to rise
again at energies $W>1.8$ GeV. However, because of the energy
weighting, the effect is negligible for the integrals.

\section{Integrals}\label{sec:int}
\subsection{Results for Real Photons}
In Tab.~\ref{tab1} we show our results for the GDH integral and the
forward spin polarizability over the lab energy range of 200-450 MeV
together with the latest Mainz data\cite{ahre00}. In comparison with
the results of the dispersion theoretical partial wave analysis
HDT\cite{HDT} and the SAID solution SM99K\cite{SAID} our MAID results
agree very well with the experiment. The additional information on
the individual channels, however, offers interesting insights in the
different calculations, especially the $\pi^+$ result of SAID for
$\gamma_0$ is a factor 3 larger than MAID and about a factor 2 above
the data. The reason is the enhanced sensitivity of background
contributions in the $\pi^+$ channel, especially the S-wave near
threshold.
\begin{table}[htbp]
\begin{center}
\caption{\label{tab1} Contributions to the  GDH integral
$I=\int_{200}^{450}(\sigma_{1/2}-\sigma_{3/2})/\nu d\nu$ and to
the forward spin polarizability of the proton
$\gamma_0=1/4\pi^2\int_{200}^{450}(\sigma_{1/2}-\sigma_{3/2})/\nu^3
d\nu$ for photon $lab$ energies of 200-450 MeV.} \vskip 0.0 cm
\begin{tabular}{ccccc}
\hline
 $I (\mu b)$ & Mainz exp.\cite{ahre00}  & SM99K\cite{SAID} &
 HDT\cite{HDT} & MAID\cite{Dre98a} \\ \hline
 $\pi^0 p$ & -124 $\pm$ 11 & -132 & -144 & -136 \\
 $\pi^+ n$ &  -33 $\pm$ 3  &  -55 &  -26 &  -23 \\
 total     & -157 $\pm$ 11 & -187 & -170 & -159 \\
\hline
 $\gamma_0 (10^{-4}fm^4)$ &   &  &  &  \\ \hline
 $\pi^0 p$ & -1.2 $\pm$ 0.3 & -1.34 & -1.48 & -1.40 \\
 $\pi^+ n$ & -0.23$\pm$ 0.04& -0.54 & -0.19 & -0.17 \\
 total     & -1.4 $\pm$ 0.3 & -1.88 & -1.67 & -1.57 \\
\hline
\end{tabular}
\end{center}
\end{table}
\begin{table}[htbp]
\begin{center}
\caption{\label{tab2} Contributions to the GDH integral for proton
and neutron: Sum rules $-2\pi^2\alpha\kappa_N/m^2$ (sr), Mainz
experiment\protect\cite{Ped00} in the energy interval of 200-800 MeV,
MAID $1\pi$ contributions, eta production\protect\cite{Kno95},
reggeized $2\pi$ contributions (Born terms and $D_{13}(1520)$
resonance)\protect\cite{holv00}.}
\begin{tabular}{ccccccccc}
\hline
 $I (\mu b)$ & sr & exp. &
 $\gamma,\pi^0$ & $\gamma,\pi^\pm$ &
 $\gamma,\eta$ & $\gamma,\pi\pi\, B$ &
 $\gamma,\pi\pi\, D $& sum \\
 \hline
 prot.  & -205 & -216
  $\pm$ 6 & -150 & -21 & +15 & -30 & -15 & -201 \\
 neut. & -233 &      ---     & -154 & +30 & +10 & -35 & -15 & -164 \\
\hline
\end{tabular}
\end{center}
\end{table}
Tab.~\ref{tab2} shows the GDH integral over the full energy range up
to $W_{max}=2$~GeV. The preliminary experimental result is obtained
only from measurements at Mainz and covers the energy range from 200
to 800 MeV photon $lab$ energy. However, our theoretical calculations
indicate a very close cancellation between the low energy
contribution from threshold up to 200 MeV (30 $\mu b$) and of
energies above 800 MeV (-34$\mu b$). For our detailed comparison we
included recent calculations of reggeized $\pi\pi$ photoproduction by
Holvoet and Vanderhaeghen\cite{holv00} that include
$\gamma,\pi\Delta$ Born terms and additional $D_{13}(1520)$
excitations. This $2\pi$ contribution to the GDH integral is about
twice as large as compared to our simple phenomenological multi-pion
parametrization used for finite $Q^2$.

Our calculation that also include very recent Regge-type
calculations for two-pion photoproduction\cite{holv00} shows a
very good agreement with the sum rule for the proton target but
also exhibits a big deviation for the neutron target. However, on
the neutron target the photoproduction information is rather
limited above the $\Delta$ region and it has to be further
investigated if the high-energy region is perhaps more important
than for the proton target. Furthermore, for both nucleon targets
high energy contributions beyond the two-pion production have to
be studied that make a very big contribution in deep inelastic
scattering at finite $Q^2 > 1 GeV^2$.
\begin{table}[htbp]
\begin{center}
\caption{\label{tab3} Contributions to the forward spin
polarizability for proton and neutron: Mainz
experiment\protect\cite{Ped00} in the energy interval of 200-800
MeV, experimental value plus threshold and high energy region,
MAID $1\pi$ contributions, eta plus multi-pion production. Values
are given in units of $10^{-4} fm^4$}
\begin{tabular}{ccccccccc}
\hline
 $\gamma_0$ & exp & exp +  &
 $\gamma,\pi^0$ & $\gamma,\pi^\pm$ &
 $\gamma,\eta$ + & sum \\
 \hline
 proton   & -1.71 $\pm$ 0.09 & -0.78 $\pm$ 0.09 & -1.47 & 0.80 & -0.01 & -0.68 \\
 neutron  &      ---         &       ---        & -1.50 & 1.66 & -0.01 &  0.15 \\
\hline
\end{tabular}
\end{center}
\end{table}
In Tab. \ref{tab3} we give the individual contributions to  the
forward spin polarizability $\gamma_0$ in the resonance region,
$W_{thr}<W<2$~GeV. Due to the strong energy weighting $1/\nu^3$, the
high-energy contribution beyond $\nu=800$MeV is practically
negligible. On the other side, the threshold region below
$\nu=200$MeV is strongly enhanced in comparison to the GDH integral.
To compare with the value of the Mainz experiment we have added the
low-energy contribution to the measured value and have obtained an
experimental value for the full energy range. In particular, the
correction is very much dominated by low-energy constraints,
therefore the estimate is rather model-independent.

\subsection{Results for Virtual Photons}
\begin{figure}[htbp]
\centerline{ \psfig{file=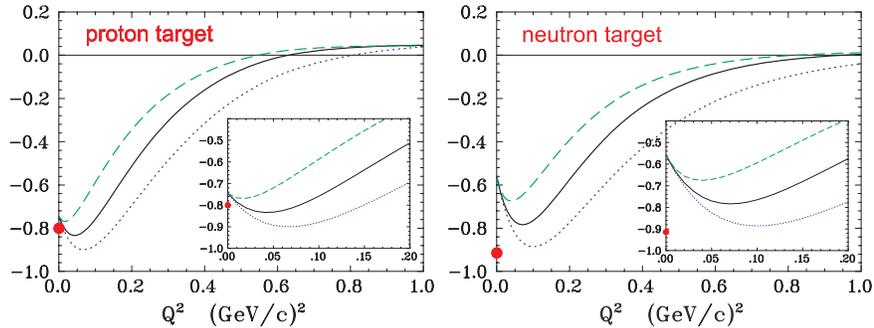,width=11.5cm,silent=}}
 \caption{\label{fig2} Generalized GDH integrals $I_{GDH}(Q^2)$ for
   3 different definitions used in the literature. The full, dashed
   and dotted lines show the integrals (a), (b), (c) in the notation
   of Eqs. \ref{eq4}, \ref{eq5}, \ref{eq6}, respectively. The integrals
   are evaluated up to $W=2$~GeV and include $1\pi + \eta +n\pi$
   contributions.}
\end{figure}

\begin{figure}[htb]
\centerline{\psfig{file=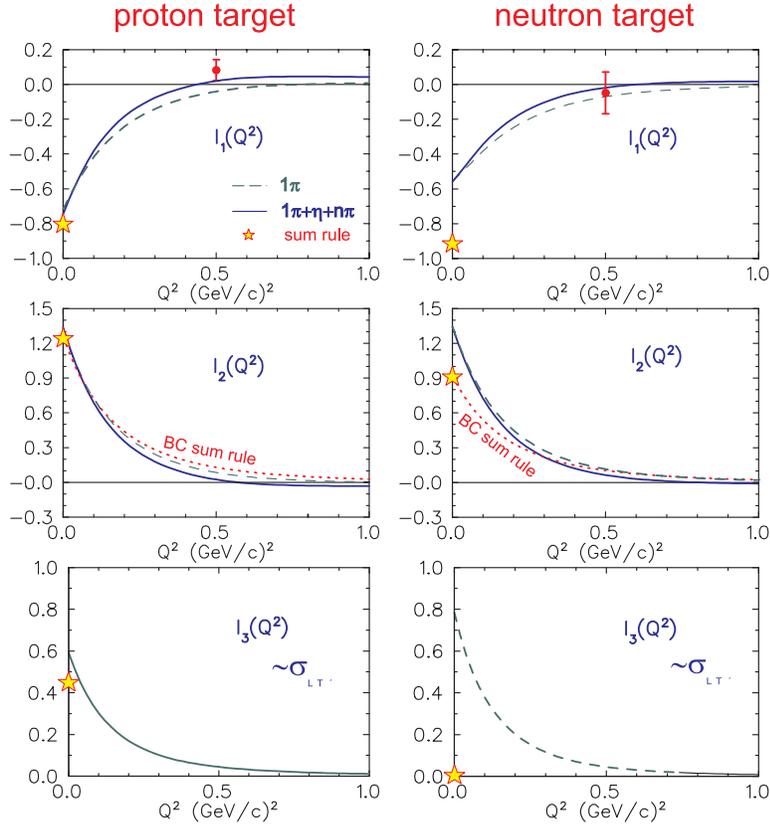,height=10.9cm,angle=0,silent=}}
\caption{\label{fig3} Generalized GDH integrals $I_{1,2,3}(Q^2)$
for the proton and the neutron integrated up to $W_{{\rm max}}=2$
GeV. The dashed lines show the contributions from the $1\pi$
channel while the full lines include $1\pi + \eta +n\pi$. The
dotted line for $I_2$ is the $BC$ sum rule prediction
of~{Ref.\protect\cite{Bur70}}. The data is from SLAC,
Ref.\protect\cite{Abe98}. The stars show the sum rules at $Q^2=0$}
\end{figure}

In Fig.~\ref{fig2} and Fig.~\ref{fig3}  we give our predictions for
the integrals $I_{GDH}(Q^2)$, $I_1(Q^2)$, $I_2(Q^2)$ and $I_3(Q^2)$
in the resonance region, i.e. integrated up to $W_{{\rm max}}=2$ GeV
for the proton and neutron targets. A comparison of the 3 different
forms, defined in Eqs. \ref{eq4}, \ref{eq5}, \ref{eq6} show
significantly different slopes at $Q^2=0$ and quite different zero
positions, where the GDH integral crosses from negative values
observed for real photons to positive values known from deep
inelastic scattering. In the case of the integral $I_1,$ our model is
able to generate the expected drastic change in the helicity
structure at low $Q^2$.
We find a zero-crossing at $Q^2=0.75$
(GeV/c)$^2$ if we include only the one-pion contribution. This
value is lowered to 0.52 (GeV/c)$^2$ and 0.45 (GeV/c)$^2$ when we
include the $\eta$ and the multi-pion contributions respectively.
The SLAC analysis of the proton yields $I_1=0.1\pm0.06$ at $Q^2=
0.5$ (GeV/c)$^2$, while our result at this point is only slightly
positive. For the neutron our calculation is fully consistent
within the SLAC analysis at $Q^2= 0.5$ (GeV/c)$^2$ in contrast to
the large discrepancy observed at $Q^2=0$, see Tab. 2.

Comparing with the generalizations of the GDH sum rule in
Fig.~\ref{fig2} and Fig.~\ref{fig3}, it can be seen that the slope at
$Q^2=0$ and the existence of a minimum for small $Q^2$ depends on the
inclusion of the longitudinal contributions, i.e. the minimum
disappears when $\sigma_{LT'}$ is added. Concerning the integral
$I_2$, our full result is in good agreement with the prediction of
the BC sum rule.  The deviation is within $10~\%$ and should be
attributed to contributions beyond $W_{\rm max}=2$ GeV and the
uncertainties in our calculation for $\sigma_{LT'}$. As seen in Eq.
(\ref{eq9}) the integral $I_3$ depends only on this $\sigma_{LT'}$
contribution. From the sum rule result a value of $e_N\kappa_N/4$ is
expected at $Q^2=0$, i.e. 0.45 for the proton and zero for the
neutron target. While our value arising entirely from the $1\pi$
channel (0.59) gets relatively close to the sum rule result for the
proton, in the neutron case this sum rule is heavily violated
(0.78). So far it is not clear where such a large negative
contribution should arise for the neutron target in order to cancel
the $1\pi$ contribution. Either it is due to the high-energy tail
that may converge rather slowly for the unweighted integral $I_3$, or
the multi-pion channels could contribute in such a way, while the eta
channel is very unlikely. In any case a careful study of the
multi-pion contribution for both proton and neutron targets will be
very helpful, in particular one can expect longitudinal contributions
from the non-resonant background.

\section{Summary}

In summary, we have applied our recently developed unitary isobar model
for pion electroproduction to calculate generalized GDH integrals and
the BC sum rule for both proton and neutron targets. Our results
indicate that both the experimental analysis and the theoretical models
have to be quite accurate in order to fully describe the helicity
structure of the cross section in the resonance region.

While our results agree quite well for the GDH and BC sum rules for the
proton, we find substantial deviations for the neutron target, in
particular the sum rule $I_3(0)\equiv I_1(0)+I_2(0)=0$ is heavily
violated by the contribution from the single-pion channel which is even
larger than in the case of the proton.
Concerning the theoretical description, the treatment of the multi-pion
channels has to be improved with more refined models. On the
experimental side, the upcoming results from measurements with
real and virtual photons from ELSA, GRAAL and JLab hold the promise to
provide new precision data in the resonance region.

\section*{Acknowledgments}
 We would like to thank M. Vanderhaeghen for the contributions to the
 $\pi\pi$ channels and J.\ Arends for the information on the experimental data analysis.
 This work was supported by the Deutsche Forschungsgemeinschaft (SFB 443).

\end{document}